\documentclass[aps,pre,floats,twocolumn,showpacs,superscriptaddress,reprint]{revtex4-1}

\usepackage{graphicx,epsfig}% Include figure files
\usepackage{graphics,dcolumn,bm,epic, eepic,fleqn,float}
\usepackage{amssymb,amsmath,multirow,rotate,color}
\usepackage{xcolor}

\begin{document}
\color{blue}

\title{Motion-induced synchronization in metapopulations of mobile agents}

%TC:ignore
\author{Jes{\'u}s G{\'o}mez-Garde\~{n}es}
\affiliation{Departamento de F\'{\i}sica de la Materia Condensada,
  University of Zaragoza, Zaragoza 50009, Spain}
\affiliation{Institute for Biocomputation and Physics of Complex
Systems (BIFI), University of Zaragoza, Zaragoza 50018, Spain}

\author{Vincenzo Nicosia}
\affiliation{School of Mathematical Sciences, Queen Mary University of London, 
London E1 4NS, United Kingdom} 
\affiliation{Computer Laboratory, University of Cambridge, Cambridge
  CB3 0FD, United Kingdom}

\author{Roberta Sinatra}
\affiliation{Center for Complex Network Research and Department of Physics, Northeastern University, Boston, Massachusetts 02115, USA}

\author{Vito Latora}
\affiliation{School of Mathematical Sciences, Queen Mary University of London, 
London E1 4NS, United Kingdom}  
\affiliation{Dipartimento di Fisica e Astronomia, Universit\`a di Catania and INFN, 95123 Catania, Italy}

\begin{abstract}
  We study the influence of motion on the emergence of synchronization
  in a metapopulation of random walkers moving on a heterogeneous
  network and subject to Kuramoto interactions at the network
  nodes. We discover a novel mechanism of transition to macroscopic
  dynamical order induced by the walkers' motion. Furthermore, we
  observe two different microscopic paths to synchronization:
  depending on the rules of the motion, either low-degree nodes or the
  hubs drive the whole system towards synchronization. We provide
  analytical arguments to understand these results.
\end{abstract}

%TC:endignore

\maketitle

\section{Introduction}

The spontaneous emergence of synchronization in systems of coupled
dynamical units~\cite{piko,boccbook} underlies the development of
coordinated tasks as diverse as metabolic cycles in eukaryote cells,
cognitive processes in the human brain and opinion formation in social
systems~\cite{strogatzrev,bullmore,socialdyn,latora_opinion}. In the last
decade complex networks theory has revealed that the topology of the
interactions in a complex system has important effects on its
collective behavior~\cite{strogatz,rev:bocc}. As a consequence, many
recent studies have considered dynamical systems coupled through
non-trivial topologies~\cite{PRsync}, uncovering the impact of the
structure of the network on the
existence~\cite{Pacheco,Arenas,gardenes,lodato,explosive} and
stability~\cite{nishikawa,chavez,zhou,delosrios} of synchronized states.

Quite frequently, the interactions among the 
units of a complex system keep changing over time. Their evolution can
either be driven by the synchronization process itself, as in models
of coevolving networks~\cite{ZhouPRL06,OttPRL08,AokiPRL09,BoccPRL11}, or be
determined by the fact that each unit moves at random over a
continuous and homogeneous space and interacts only with other units
within a given distance~\cite{buscarino,frasca1,diaz,frasca2,diaz2}.

In many cases, the motion of the agents takes place on discrete and
heterogeneous media, that can be represented as complex
networks. Typical examples include users browsing the World Wide Web,
airplane passengers traveling throughout a country, or people playing
online social games~\cite{usair,vittoria3,szell}.  In such systems,
both the rule of motion adopted by the agents, and the heterogeneity
of the environment, have an impact on the emergence and stability of
collective behaviors.
%%%
%%However, it is difficult 
%  to include these factors in solvable analytical
%  models. 
For this reason, metapopulation modeling has been successfully
employed to explore the combined effect of mobility and non-trivial
interaction patterns in different contexts, including the study of
epidemic spreading and chemical
reactions~\cite{usair,vittoria3,vittoria1,vittoria2,meloni1,meloni2}.

In this work we propose a metapopulation model to study the
emergence of synchronization in populations of individuals moving over
discrete heterogeneous environments, and interacting through nonlinear
dynamical equations. We assume that each agent is characterized by an
internal state (or opinion), and that it moves over the environment
trying to synchronize its opinion with that of the other
individuals. Thus, the evolution of the system is driven by the
interplay of two concurrent processes: on one hand, the interaction of
neighboring agents drives their internal state towards local
consensus; on the other hand, agents' motion dynamically changes the
pattern of interaction and allows each agent to be exposed to
different opinions.  We discover a novel mechanism of synchronization
that we name {\em motion-induced synchronization}, since the
transition from disorder to macroscopic order is controlled by the
value of the parameter tuning the motion of agents. Furthermore, we
show that there are two different microscopic mechanisms driving the
system towards synchronization, according to whether the walkers
prefer to visit or to avoid high-degree nodes.

\section{The model}

Our metapopulation model consists of two layers. At the bottom layer
we have a set of $W$ mobile agents (walkers). Each agent $i$
($i=1,2,\ldots W$) is a dynamical system whose internal state at time
$t$ is described by a phase variable, $\theta_i(t)\in
\left[0,2\pi\right)$, %
and changes over time as a result of the interactions with other
agents. The top layer consists of a complex network with $\text{N}$
nodes and $\text{E}$ edges, which represents the environment into
which agents interact (nodes) and move (edges). The network is
described by an adjacency matrix $\mathcal{A}$, whose entry $a_{IJ}$
is equal to $1$ if nodes $I$ and $J$ are connected by an edge, and $0$
otherwise (here and in the following we indicate nodes of the graph in
uppercase letters, and walkers in lowercase). 

At any given time, each
agent is located in one of the nodes of the network. The agent
interacts for a fixed time interval with other agents at the same
node, trying to synchronize its phase with the others'. Then, it moves
to one of the neighboring nodes, chosen according to a one-parameter
motion rule.
More precisely, assume that at time $t$ we have $i\in I$, {\em i.e.}
agent $i$ is at node $I$. The evolution of the phase $\theta_i(t)$ of
agent $i$ is ruled by an all--to--all Kuramoto-like interaction with
the other walkers being on the same node $I$ at time
$t$~\cite{kuramoto,strogatzrev,acebron}:
\begin{equation}
 \dot{\theta}_i(t)=\omega_i+\lambda\sum_{j\in
   I}\sin(\theta_j(t)-\theta_i(t))\;,~\forall i \in I\;
 %%I=1,\ldots,\text{N}
\label{kura}
\end{equation}
where $\omega_i$ is the internal frequency of agent $i$ and $\lambda$
is a control parameter accounting for the strength of the interaction
among walkers.  
Notice that, when the phases of the agents evolve according to
Eq.~(\ref{kura}), the system is not driven by a single global
mean-field (as in the classical all--to--all Kuramoto model). Instead,
each oscillator $i$ interacts with the \textit{local} mean-field phase
due to all the oscillators being at the same node as $i$.

At regular time intervals of length $\Delta$ all the agents perform
one step of a \textit{degree-biased} random walk on the network.
Namely, we assume that a walker at node $I$ moves to a neighboring
node $J$ with a probability proportional to the degree $k_J$ of the
destination node~\cite{pre,sinatra}:
\begin{equation}
\Pi_{I\rightarrow J} =
   \frac{a_{IJ} k_{J}^{\alpha}}{\sum_{L=1}^{\text{N}} a_{IL}k_{L}^{\alpha}}\;.
\label{rw}
\end{equation}
Here $\alpha$ is a tunable parameter which biases agents' motion
either towards low-degree nodes ($\alpha<0$) or towards hubs
($\alpha>0$). For $\alpha=0$, we recover the standard (unbiased)
random walk. 
In summary, the metapopulation model has three control parameters:
$\lambda$ regulating the interaction strength among walkers, $\alpha$
tuning the rule of their motion, and $\Delta$ fixing the ratio between
the time scales of interaction and motion. 
%
%

%  ORDER PARAMETERS
The degree of synchronization of the whole metapopulation at time $t$
is measured by the global order parameter:
\begin{equation}
r(t)=\left|\frac{1}{W}\sum_{i=1}^{W}{\mbox e}^{{\mbox
    i}\theta_i(t)}\right|\;
\label{rglobal}
\end{equation}
where $r\simeq 0$ if the phases of the agents are completely
incoherent, while $r=1$ when the system is fully synchronized. In
order to quantify the degree of synchronization of a single node $I$
we introduce the local order parameter:
\begin{equation}
r_I(t) = \left|\frac{1}{w_I(t)}\sum_{i \in I}{\mbox e}^{{\mbox
    i}\theta_i(t)}\right|\;, ~~~I=1,2,\ldots,\text{N}
\label{rlocal}
\end{equation}
where $w_{I}(t)$ is the number of agents at node $I$ at time $t$.
When the phases of the walkers at node $I$ are fully synchronized, the
local order parameter of the node is equal to 1, while in the case of
complete local disorder we have $r_I(t)=0$.  We can also quantify the
average local synchronization of the network $r_{\text{loc}}(t)$ as
the average of $r_I(t)$ over all nodes, i.e.:
\begin{equation}
   r_{\text{loc}}(t)  = \frac{1}{N}\sum_I r_I(t)
  \label{ravg}
\end{equation}
We notice that, having $r_I(t)\simeq 1\;\forall I$, or equivalently,
$r_{\text{loc}}(t)\simeq 1$ is a necessary but not sufficient
condition to attain global synchronization. In fact, in the limiting
case in which there is no motion ($\Delta\rightarrow\infty$) and
$\lambda$ is large enough, it is possible to have $r_I(t)\simeq
1\;\forall I$ and, at the same time, $r(t)\simeq 0$.

\section{Results}

We have simulated the metapopulation model on various synthetic
networks and, as an example of a real complex network, on the US
air-transportation system, which includes the flight connections,
between the $\text{N}=500$ largest airports in the US.  Thanks to its
intrinsic nature as a backbone for human transportation, the US
airports network has already been used to investigate
reaction-diffusion dynamics in metapopulation
models~\cite{usair}. This network has a long-tailed degree
distribution, exhibits degree-degree correlations and is relevant for
the present study because opinion formation in real social systems is
often mediated by information, communication and transportation
networks having similar structural properties. We have generated, for comparison, uncorrelated scale-free (SF)
networks with $\text{N}=500$ nodes and a power-law degree distribution
$P(k)\sim k^{-\gamma}$ with a tunable value of the exponent
$\gamma$~\cite{configuration}. 

For the numerical simulations of the model, the initial phases of the
oscillators have been sampled uniformly in $[0,2 \pi)$, and their
internal frequencies $\omega_i$ from a uniform distribution
$g(\omega)=1/2\;\forall\;\omega\in[-1,1]$. We started from a
stationary distribution of $W=5000$ walkers over the
networks~\cite{pre}, and we integrated Eq.~(\ref{kura}) for all the
agents for a time $t_0=m\Delta$, where $m=10^4$ is the number of
random walk steps performed. After this transient, we estimated
global and local synchronization parameters ($r$, $r_{\text{loc}}$,
and $r_I$ for $I=1,\ldots,\text{N}$) by respectively averaging the
values obtained from Eq.~(\ref{rglobal}), (\ref{rlocal}) and (\ref{ravg}) over a time window of length $T=2m\Delta$.

\begin{figure}[t!]
\centering
\vspace{-0.3cm}
\includegraphics[width=3in,angle=-0]{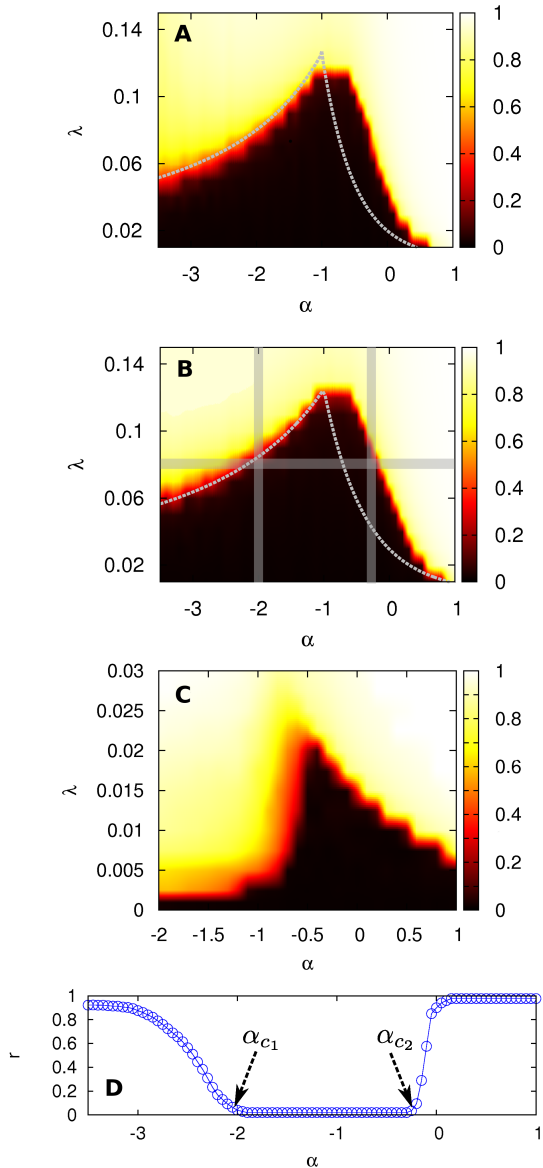}
\caption{(color online) Phase diagram $r(\alpha,\lambda)$ of the
  metapopulation model on uncorrelated SF networks with $\gamma=2.7$
  (panel A), $\gamma=3.0$ (panel B) and on the US air-transportation network (panel C). The three
  networks have $\text{N}=500$ nodes. The dashed curves in panels A and B are the corresponding lower-bound 
  analytical predictions for the onset of synchronization in uncorrelated graphs obtained from  Eq.~(\ref{low1}) for $\alpha>-1$, and from Eq.~(\ref{low2}) for  $\alpha<-1$. Panel D shows $r(\alpha)$ for $\lambda=0.08$,
  corresponding to the horizontal line in panel B.}
\label{fig1}
\end{figure}

\subsection{Synchronization transition}
In Fig.~\ref{fig1} we show the global order parameter $r$ as a
function of the coupling strength $\lambda$, and of the walker bias
$\alpha$. The three phase diagrams have been obtained setting
$\Delta=0.05$, but qualitatively similar results have been obtained
for different values of $\Delta$. The SF networks reported in
Fig.~\ref{fig1}.A and B have respectively $\gamma=2.7$ and $\gamma=3$.
As expected, by increasing $\lambda$ at a fixed value of $\alpha$,
{\em i.e.} keeping fixed the rules of motion, we observe a phase
transition from the incoherent phase ($r\simeq0$, dark regions of the
diagrams) to a synchronized state ($r\neq 0$, bright regions of the
diagrams). However, the precise value for the onset of
synchronization, namely the critical value $\lambda_c$ for which the
incoherent state becomes unstable, strongly depends on the motion bias
$\alpha$. In particular, we find that $\lambda_c(\alpha)$ is first
increasing as function of $\alpha$, and then decreasing. The function
$\lambda_c(\alpha)$ reaches its maximum $\lambda^{\text{max}}_c$ at a
particular value of $\alpha$, namely at $\alpha^*\simeq-0.5$ for the
air-transportation network, and at $\alpha^*\simeq -1$ for the two SF
networks. By comparing the diagrams obtained for the two SF networks
we also observe that, for any value of $\alpha$, the critical value of
$\lambda_c$ is smaller for SF networks with $\gamma=2.7$ (panel A)
than for those having $\gamma=3.0$ (panel B), thus confirming that
degree heterogeneity tends to favor global synchronization.

As a consequence of the shape of $\lambda_c(\alpha)$, for a wide range
of values of the strength $\lambda$ such that
$\lambda<\lambda^{\text{max}}_c$, we observe a novel mechanism of {\em
  motion-induced} synchronization. This means that we can fix the
value of the interaction strength $\lambda$, and we can control
whether the system is in the incoherent phase ($r\simeq0$, dark
regions) or in the synchronized state ($r\neq 0$, bright regions)
solely by changing the rule of motion.

Let us focus on the case of the SF network shown in Fig.~\ref{fig1}.B
and suppose to keep $\lambda$ fixed at $0.08$ (see the horizontal line
in the figure). As shown in Fig.~\ref{fig1}.D, we can start with a
value of $\alpha$ within the incoherent phase, for instance
$\alpha=\alpha^{*}=-1$, and consider the behavior of the system as we
decrease the value of $\alpha$. When $\alpha$ gets smaller than a
particular critical value $\alpha_{c_1}(\lambda)$, in this case
$\alpha_{c_1}(0.08)\simeq -2.0$, we observe a transition from the
incoherent to the synchronized phase. Conversely, we can start at
$\alpha=-1$ and get a synchronized state by increasing the motion bias
parameter to values larger than $\alpha_{c_2}(0.08)\simeq -0.2$. A
fine tuning of the rules controlling the agents' motion can
effectively produce dramatic changes in the \textit{macroscopic}
synchronization state of the system. 

We have found that the bias in the motion affects the onset of
synchronization also at the \textit{microscopic} level. To illustrate
this result we look at the microscopic paths to
synchronization~\cite{gardenes} as we increase $\lambda$, by following
the two vertical lines shown in Fig.~\ref{fig1}.B. In particular, we
have computed the value of the local order parameter for each of the
nodes of the graph, as in Eq.~(\ref{rlocal}), and we have grouped
nodes by degree classes. We computed the average value of the local
synchronization of nodes of degree $k$ as:
\begin{equation} 
r_k= \frac{1}{N_k} \sum_{I=1}^{\text{N}} r_I\delta(k_I,k)\;
\label{eq:rk}
\end{equation} 
where $N_k=\text{N}P(k)$ is the number of nodes of degree $k$. 

\begin{figure}[t!]
\centering
\includegraphics[width=3.25in]{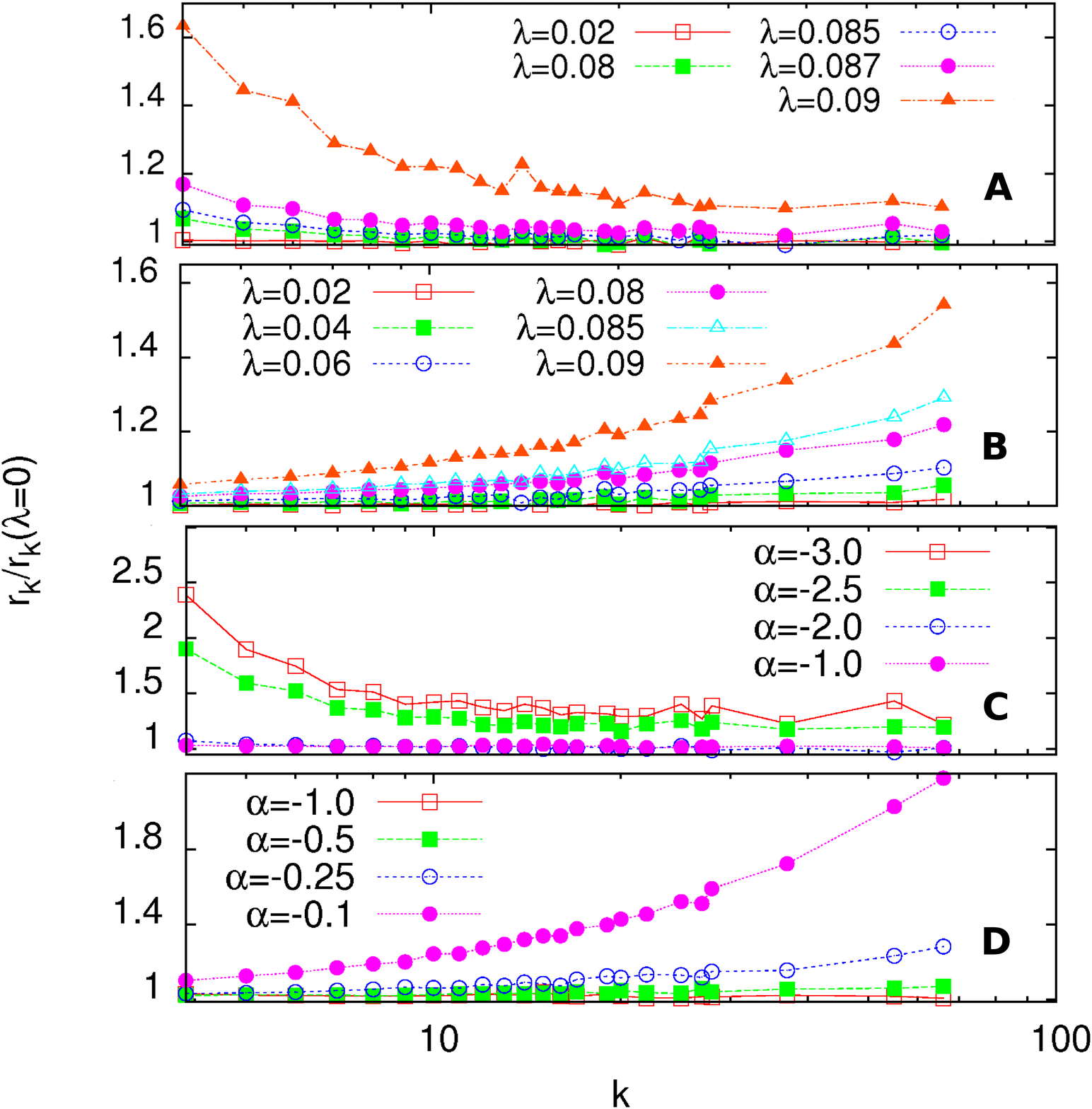}
\caption{(color online) Panel A and B: the average local order
  parameter $r_k$ of nodes of degree $k$ as a function of $k$, for
  various values of $\lambda$, and for two fixed values of the bias,
  respectively $\alpha=-2.0$ (panel A) and $\alpha=-0.25$ (panel B),
  corresponding to the two vertical lines in Fig.~\ref{fig1}.B. Panel
  C and D: $r_k$ for $\lambda=0.08$ and various values of $\alpha$
  (respectively, $\alpha<-1$ in panel C and $\alpha>-1$ in panel D)
  corresponding to the horizontal line in Fig.~\ref{fig1}.B.}
\label{fig2}
\end{figure}

We report in Fig.~\ref{fig2}.A and \ref{fig2}.B the quantity $r_k$
divided by the value $r_k(\lambda\simeq 0)$ obtained when $\lambda$ is
close to $0$ as a function of $k$, and for different values of
$\lambda$. Panel A corresponds to $\alpha=-2.0$ and panel B to
$\alpha=-0.25$. When $\alpha=-2.0$ the nodes having small degree are
the first ones to attain local synchronization as soon as $\lambda$
crosses the critical value $\lambda_c(-2.0)\simeq 0.08$; conversely,
for $\alpha=-0.25$, the hubs are the nodes which synchronize first
when $\lambda>\lambda_c(-0.25)\simeq 0.07$.  We thus observe two
microscopic paths to synchronization: either driven by low-degree
nodes ($\alpha<\alpha^*$), or by the hubs ($\alpha>\alpha^*$). The two
different synchronization mechanisms are also evident by following the
horizontal line in Fig.~\ref{fig1}.B, i.e. by plotting $r_k$ for a
fixed value of $\lambda$ and different values of $\alpha$, as shown in
Fig.~\ref{fig2}.C and \ref{fig2}.D.

\subsection{Analytical estimation of the synchronization threshold}

The effects of motion on synchronization can be explained by
analytical arguments in the case of networks without degree-degree
correlations. In particular we derive, as follows, a lower-bound
estimate for the critical strength $\lambda_c$ as a function of
$\alpha$.
The average number $w_I$ of biased random walkers at a node $I$ of an
undirected connected graph without degree-degree correlations
reads~\cite{pre,sinatra}:
\begin{equation}
w_I=\frac{W c_Ik_I^{\alpha}}{\sum_{J=1}^{\text{N}}c_Jk_J^{\alpha}}
  \simeq \frac{W k_I^{\alpha+1}}{\sum_{J=1}^{\text{N}}k_J^{\alpha+1}}
  =\frac{ W k_I^{\alpha+1}}{\text{N}
  \langle k^{\alpha+1} \rangle}\;,
\label{stat}
\end{equation}
where $c_I=\sum_{J=1}^{\text{N}}a_{IJ}k_J^{\alpha}$. For a given
$\alpha$, the value $w_I$ depends only on the connectivity $k_I$ of
the node, so that all the nodes with the same degree will have the
same average number of walkers. Thus, in the following we indicate as
$w_k$ the number of agents on a node with degree $k$.

We consider now the two limiting cases $\Delta\rightarrow 0$ and
$\Delta\rightarrow\infty$. When $\Delta\rightarrow 0$ (fast-switching
approximation) the agents on a node interact for an infinitesimal time
interval before moving to another node. In this limit we have a
well-mixed population of oscillators that can be approximated as a
single all-to-all Kuramoto model of $W$ elements. Thus, the critical
value of the coupling $\lambda$, in this case reads
$\Lambda_c=2/[W\cdot\pi\cdot g(0)]$~\cite{strogatzrev,acebron}, and
does not depend on $\alpha$. 

When $\Delta\rightarrow\infty$ (slow-switching approximation), i.e. when the walk is much slower than
the Kuramoto dynamics, each node of the network is an all-to-all
Kuramoto system independent from the others. For a fixed value of
$\lambda$, in some nodes the oscillators will reach local
synchronization before eventually walking away, while in some other
nodes they will not. In fact, the critical coupling strength of a node
$I$ is that of a set of $w_I$ all-to-all coupled Kuramoto oscillators:
$\lambda_c(I)=2/[w_I\pi\cdot g(0)]$. Hence, the critical coupling
strength for the local synchronization of the walkers at a node of
degree $k$ reads:
\begin{equation}
\lambda_c(k)=\frac{2}{w_k\pi g(0)}=\frac{4\text{N} \langle
  k^{\alpha+1} \rangle}{W\pi k^{\alpha+1}}\;,
\label{lambdai}
\end{equation}
where we have made use of Eq.~(\ref{stat}). 
Therefore, in the slow-switching approximation, at fixed values of
$W/\text{N}$, $\alpha$ and $\lambda$, only agents at nodes of degree
$k$ such that $\lambda_c(k)<\lambda$ will attain local
synchronization. Consequently, a necessary but not sufficient
condition to have global synchronization is that there is at least one
node $J$ in the graph for which $\lambda_c(k_J)<\lambda$.

Equation ~(\ref{lambdai}) sheds light on the two different microscopic paths
to synchronization observed in Fig.~\ref{fig2}. Let us indicate as
$k_{\text{min}}$ and $k_{\text{max}}$ respectively the minimum and the
maximum degree in the network. Consider two values of $\alpha$, one
larger and one smaller than $\alpha^*=-1$, for instance the two values
$\alpha=-2$ and $\alpha=-0.25$
corresponding to the two vertical lines in Fig.~\ref{fig1}.B. If we
start increasing $\lambda$ from $\lambda=0$, the slow-switching
approximation predicts no local synchronization until $\lambda$
becomes larger than the smallest value of $\lambda_c(k)$,
corresponding to $k=k_{\text{min}}$ if $\alpha<-1$, or to
$k=k_{\text{max}}$ if $\alpha>-1$.  At this point, if $\alpha<-1$
($\alpha>-1$) the walkers at nodes with the smallest (largest) degree
attain local synchronization. If we keep increasing $\lambda$, local
synchronization is progressively reached also at nodes with larger
(resp. smaller) degrees when $\alpha<-1$ (resp. $\alpha<-1$).
We can therefore derive a lower-bound ${\tilde\lambda_c}(\alpha)$ for
the curve $\lambda_c(\alpha)$ delimiting the synchronization region
%% --- REMOVED --- ($r\neq 0$)
in Fig.~\ref{fig1}.B, by considering the smallest value of $\lambda$
at which at least one class of nodes 
%% --- REMOVED --- (those having minimal or maximal degree)
attains local synchronization.
In particular, for a finite-size SF network with $P(k)\sim
k^{-\gamma}$ with $\gamma\in (2,3]$, as the ones used in our
simulations, we get:
\begin{equation}
\tilde{\lambda}_c(\alpha)=\frac{4(\gamma-1)\left[(k_{\text{min}})^{\gamma-1}-(k_{\text{min}})^{\alpha+1}(k_{\text{max}})^{\gamma-\alpha-2}\right]}{\pi
  W(\alpha+2-\gamma)}\;,
\label{low1}
\end{equation}
when $\alpha>-1$, and:
\begin{equation}
\tilde{\lambda}_c(\alpha)
=\frac{4(\gamma-1)\left[(k_{\text{max}})^{\alpha+1}(k_{\text{min}})^{\gamma-\alpha-2}-(k_{\text{max}})^{\gamma-1}\right]}
{\pi W(\alpha+2-\gamma)}\;,
\label{low2}
\end{equation}
for $\alpha<-1$. 

\begin{figure}[!ht]
    \begin{center}
      \includegraphics[width=3in]{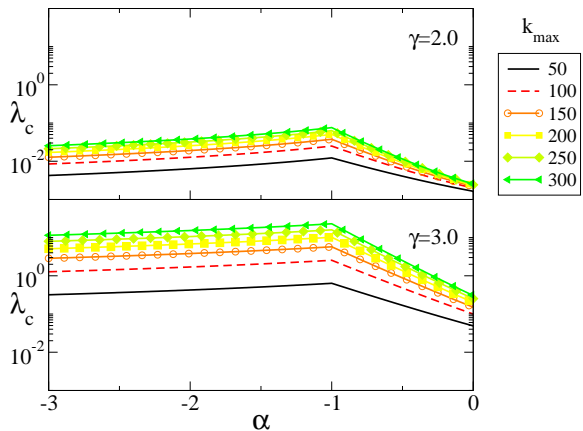}
    \end{center}
    \caption{(color online) Theoretical predictions for the critical
      coupling strength $\lambda_c(\alpha)$ as a function of $k_{max}$
      for scale-free random graphs with exponent $\gamma=2.0$ (top
      panel) and $\gamma=3.0$ (bottom panel). When $k_{max}$
      increases, global synchronization is attained for larger values
      of $\lambda$. Also an increase of $\gamma$, which corresponds to
      more homogeneous degree distributions, produces an increase of
      the critical coupling strength.}
    \label{fig:k_max}
  \end{figure}

We notice that if the graph is uncorrelated, a motion rule with
$\alpha=-1$, leads to a uniform distribution of the walkers over the
nodes, since $w_I=W/N$ $\forall I$ in Eq.~(\ref{stat}), and all the
nodes attain local synchronization altogether at
$\lambda=4\text{N}/(W\pi)$, as can be seen from Eq.~(\ref{lambdai}).
This corresponds to the largest possible value
$\lambda_c^{\text{max}}$ of the critical interaction strength.  In the
panels A and B of Fig.~\ref{fig1} we report as dashed line the curves
$\tilde{\lambda}_c(\alpha)$ obtained for the same values of
$W/\text{N}$, $k_{\text{min}}$, $k_{\text{max}}$ and $\gamma$ used in
the numerical simulations.  Although the slow-switching approximation
provides only a lower-bound for the critical interaction strength, it
works quite well for both kinds of SF networks, and it also predicts
quite accurately the position of the cusp at $\alpha=\alpha^*=-1$ for
any value of $\gamma$ in $(2,3]$. In general, Eq.~(\ref{low1}) and
Eq.~(\ref{low2}) depend on the actual value of $k_{min}$ and
$k_{max}$. However, power--law degree distributions with $\gamma\in
(2,3]$ are characterized by unbound fluctuations, so that the value of
  $k_{max}$ for scale--free random graphs having the same values of
  $N$ and $\gamma$ can vary in a substantial manner across different
  realizations. In Fig.~\ref{fig:k_max} we report the theoretical
  predictions for $\lambda_c(\alpha)$ as a function of $k_{max}$
  obtained by Eq.~(\ref{low1}) and (\ref{low2}) for two values of
  $\gamma$, namely $\gamma=2.0$ and $\gamma=3.0$. We observe that for
  fixed $\gamma$ larger values of $k_{max}$ correspond to higher
  values of the critical coupling $\lambda_c(\alpha)$. Moreover, by
  increasing $\gamma$, i.e. by moving towards more homogeneous degree
  distributions, the critical coupling for the onset of
  synchronization becomes larger. This observation confirms that
  degree heterogeity tends to promote global synchronization, as
  suggested by the phase diagrams reported in Panel A and B of
  Fig.~\ref{fig1}.
  
%%%% Importance of motion
%%%
%
%%%%
%%%%
%%%%

\begin{figure}[t!]
  \centering
  \includegraphics[width=3.in]{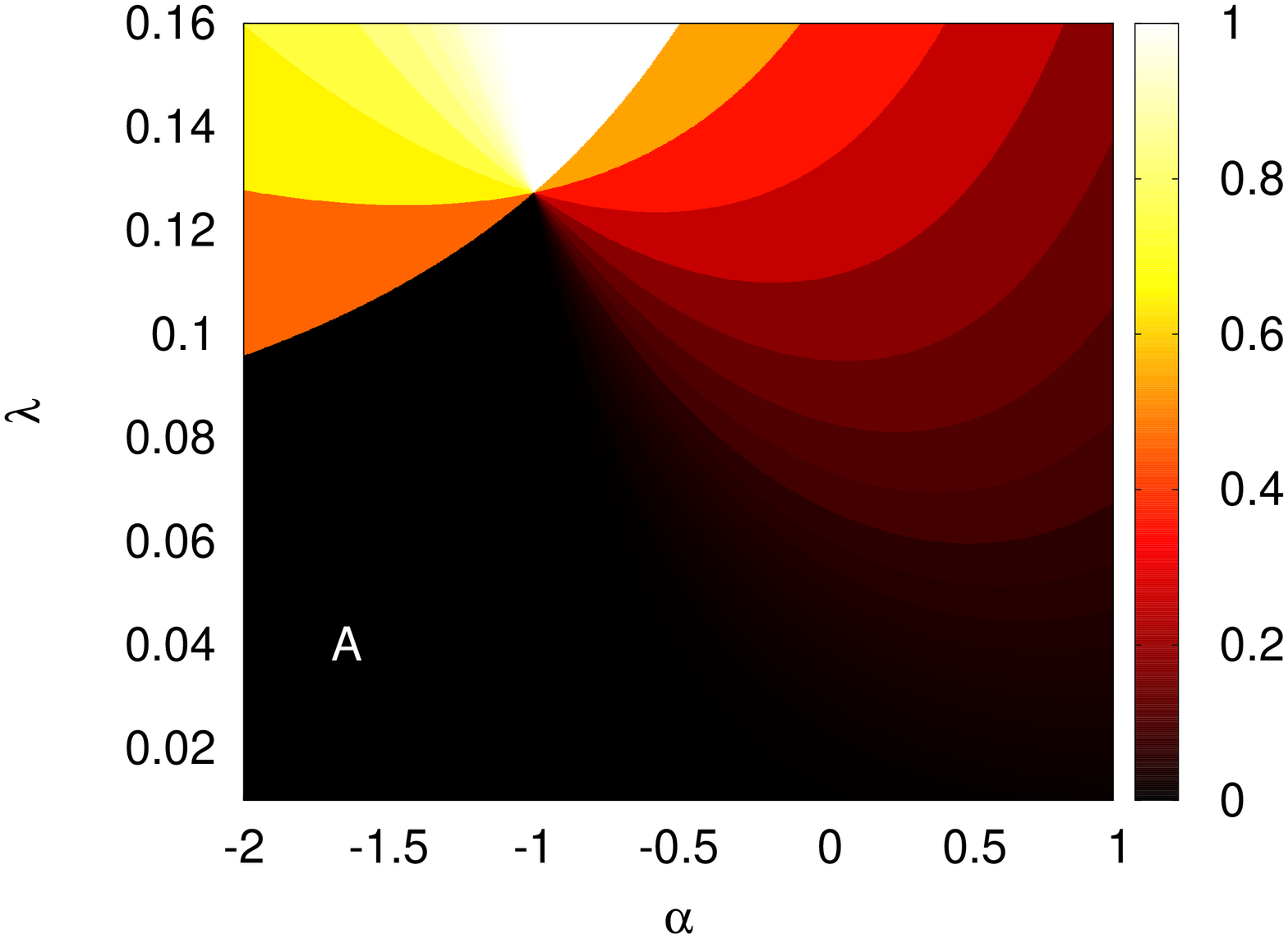}\\
  \includegraphics[width=3.in]{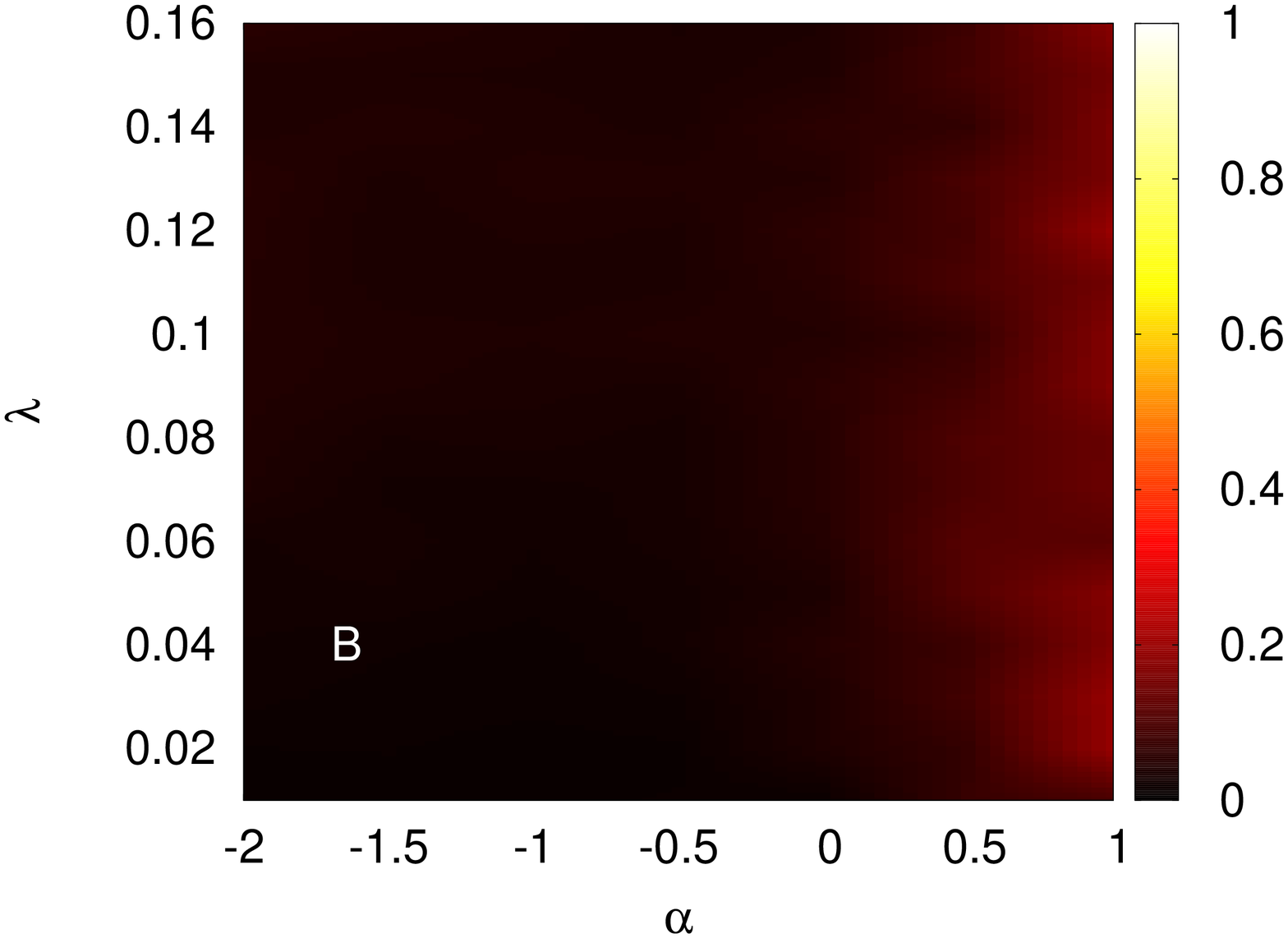}
  \caption{(color online). Phase diagram reporting the local and global
    order parameters, $r_{\text{loc}}$ (upper panel) and $r$ (lower
    panel), as a function of $\alpha$ and $\lambda$ for the
    metapopulation model in the limit $\Delta \rightarrow \infty$
    ($W=5000$ agents on a scale--free network with $N=500$
    nodes). When there is no motion, the system is globally
    incoherent, even if the local synchronization at the nodes can be
    enhanced at will by increasing the value of the interaction
    strength $\lambda$.}
  \label{fig:suppl1}
\end{figure}

\subsection{Interaction {\em vs.} Motion time scales}

We now briefly discuss the impact on synchronization of the parameter
$\Delta$, which controls how often the agents perform a step of random
walk. We first consider the model in the limiting case $\Delta
\rightarrow \infty$, in which the agents are not allowed to move.  For
each value of $\alpha$, we distributed a population of $W=5000$
walkers across the nodes of a random scale--free network with $N=500$
nodes and $P(k)\sim k^{-3}$, according to the stationary distribution
of Eq.~(\ref{stat}). Since there is no motion, each oscillator will
remain at the initial node and will interact with the same set of
oscillators for all the duration of the simulation. In this limit, the
metapopulation model is equivalent to a set of $N$ independent
all-to-all Kuramoto systems, with the system at a node of degree $k$
having a critical interaction strength given in Eq.~(\ref{lambdai}).

In Fig.~\ref{fig:suppl1} we report the local and global order
parameters, $r_{\text{loc}}$ and $r$, as a function of $\alpha$ and
$\lambda$.  We observe that, for each value of $\alpha$, there exists
a critical value of $\lambda$ such that at least one node of the
network can attain local synchronization, and by increasing $\lambda$
we can reach high values of $r_{\text{loc}}$ (panel A). Conversely,
the global order parameter $r$ always remains close to zero, and no
global synchronized state is found for any value of $\alpha$ and
$\lambda$ (panel B). Notice that the phase diagram of
Fig.~\ref{fig:suppl1}.B looks quite different from the one reported in
Fig.~\ref{fig1}A, which corresponds to a simulation with $\Delta=0.05$
on the same network. This indicates that, in the absence of motion,
the system will remain incoherent at a global scale, even if
synchronization can emerge at the level of network nodes.

\begin{figure}[!t]
  \centering
  \includegraphics[width=3in]{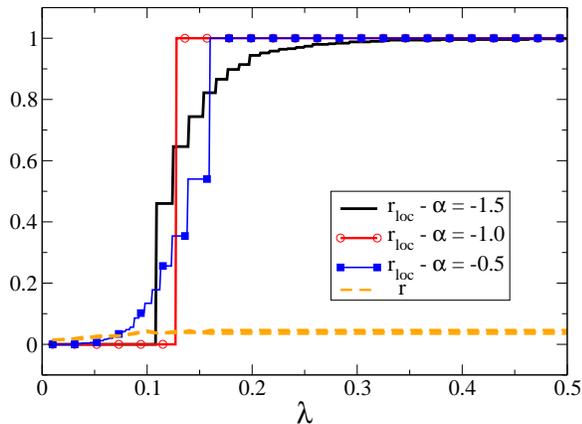}
  \caption{(color online). Cross-section plots of the two phase
    diagrams in Fig.~\ref{fig:suppl1}. We report $r_{\text{loc}}$ and
    $r$ as a function of the coupling strength $\lambda$, for three
    different values of $\alpha$.  The absence of motion hinders
    global synchronization ($r\simeq 0$, orange dashed lines), even
    if, for an appropriately large value of $\lambda$, all the nodes
    of the network can reach complete local synchronization
    ($r_{\text{loc}} =1$, solid lines).}
  \label{fig:suppl2}
\end{figure}

The behavior of the model for $\Delta \rightarrow \infty$ is better
illustrated by the cross-section plots shown in
Fig.~\ref{fig:suppl2}. Here, we report the values of local and global
order parameters, $r_{\text{loc}}$ and $r$, for three different
choices of the bias, namely for $\alpha=-1.5$ (black), $\alpha=-1.0$
(red) and $\alpha=-0.5$ (blue).  Notice that if $\lambda$ is large
enough all the nodes will eventually attain local synchronization ($
r_{\text{loc}} \simeq 1$), but for any combination of $\alpha$ and
$\lambda$, the system remains globally incoherent ($r\simeq 0$). As
expected, all the nodes achieve full local synchronization altogether
when $\alpha=-1.0$, i.e. when the system is initialized with an equal
number of walkers at each node.

\begin{figure}[!t]
  \centering \includegraphics[width=2.2in,angle=-90]{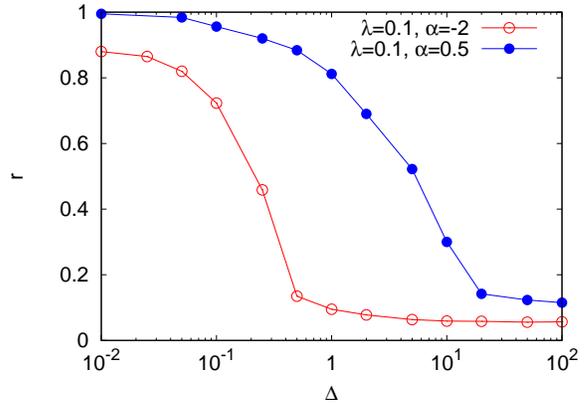}
  \caption{(color online). Global order parameter $r$ as a function of
    $\Delta$ for $\lambda=0.1$ and two values of the motion bias,
    respectively $\alpha=-2.0$ (red open circles) and $\alpha = 0.5$
    (blue filled circles). Global synchronization is attained when
    agents move frequently ($\Delta\rightarrow 0$), while the system
    remains in an incoherent state if the motion is too slow ($\Delta
    \rightarrow \infty$). }
  \label{fig:suppl3}
\end{figure}

We now consider the metapopulation model for finite values of $\Delta$. 
In Fig.~\ref{fig:suppl3} we report the value of the global order
parameter $r$ as a function of $\Delta$, for $\lambda=0.1$ and two
values of the motion bias, namely $\alpha = -2.0$ (red open circles)
and $\alpha=0.5$ (blue filled circles).
When the value of $\Delta$ is large, i.e. when the motion is rare with
respect to agents interaction, the global order parameter decreases
dramatically and approaches the behavior of the liming case
$\Delta\rightarrow \infty$. This means that even if the value of
$\lambda$ is large enough to guarantee that all the agents on a node
will attain full synchronization between two subsequent steps of the
random walk, the poor mixing due to rare motion prevents the emergence
of global order. Conversely, if $\Delta$ is small then the interaction
interval at each node is not large enough to allow local
synchronization; nevertheless, the presence of fast motion enhances
mixing and promotes the convergence of each oscillator towards a
global synchronized state. The good agreement between the prediction
in the slow-switching approximation and the numerical simulations
reported in Fig.~\ref{fig1} indicates that the value $\Delta=0.05$
corresponds indeed to an intermediate regime of the system in which
the motion is fast enough to allow a sufficient mixing and the
attainment of global synchronization and, at the same time, it is slow
enough to avoid full mixing, for which
$\lambda_c(\alpha)=\Lambda_c\;\forall\alpha$.

\section{Conclusions}

In this work we have shown a novel mechanism to induce and control synchronization, that is solely based
on the agents motion. To this end, we introduce and study a metapopulation model of random
walkers moving over a complex network. The agents obey to a one-parameter motion rule that
can bias the motion either towards low-degree nodes or towards hubs. Each walker is a Kuramoto
oscillator and interacts with the other oscillators present on the same node at a given time. To
our knowledge, this is the first time that synchronization has been studied in a metapopulation
model.

We have shown both numerically and analytically that: {\em (i)} the emergence of a synchronized phase is determined by the value of the motion bias, which effectively acts as a control parameter of a motion-induced phase transition; {\em (ii)} for each fixed value of the interaction strength, there are two critical values of the motion bias, so that a fall-and-rise of synchronization can be purely driven by motion; {\em (iii)} the two phase transitions are associated with two different microscopic paths to synchronization, respectively driven either by hubs or by low-degree nodes. 

Prior research has suggested that the strength and topology of interactions were the unique
elements driving the transition from an incoherent state to synchronization. Here we prove that
motion alone can control the onset of global coherence. This study paves the way towards further
investigations of the interplay between mobility and synchronization in complex systems.

\begin{acknowledgments}
J.G.G. and V.N. contributed equally to this work. This work has been
partially supported by the Spanish MINECO projects FIS2011-25167 and
FIS2012-38266-C02-01; and the European FET project MULTIPLEX (317532).
J.G.G. is supported by MINECO through the Ram\'on y Cajal
program. V.N. acknowledges support from the EPSRC project MOLTEN
(EP/l017321/1) and from the EU-LASAGNE Project, Contract No.318132
(STREP).  R.S.  acknowledges support from the James S. McDonnell
Foundation.
\end{acknowledgments}

%TC:endignore

\end{document}